\documentclass[manuscript]{emulateapj}
\usepackage{natbib, hyperref}
\usepackage{amsmath}
\usepackage{color}
\slugcomment{To appear in The Astronomical Journal}

\shorttitle{Lick-Index Calibration of GMOS}
\shortauthors{Puzia et al.}

\begin{document}
\title{The Lick-Index Calibration of the Gemini Multi-Object Spectrographs}

\author{Thomas H. Puzia}
\affil{Department of Astronomy and Astrophysics, Pontificia Universidad Cat\'{o}lica de Chile, Avenida Vicu\~{n}a Mackenna 4860, Macul, Santiago, Chile, {\it tpuzia@astro.puc.cl}}

\author{Bryan W. Miller}
\author{Gelys Trancho\altaffilmark{1}}
\affil{Gemini Observatory, Casilla 603, La Serena, Chile, {\it bmiller@gemini.edu}}

\author{Brett Basarab\altaffilmark{2}}
\affil{Middlebury College, Middlebury, VT 05753, USA}

\author{Jordan T. Mirocha\altaffilmark{2}}
\affil{Center for Astrophysics and Space Astronomy, University of Colorado, 389 UCB, Boulder, CO 80309, USA}

\and

\author{Karen Butler\altaffilmark{3}}
\affil{National Optical Astronomy Observatory, 950 N. Cherry Ave., Tucson, AZ 85719, USA}

\altaffiltext{1}{G.T.'s present address: GMTO Corporation, P.O. Box 90933, Pasadena, CA 91109}
\altaffiltext{2}{AURA/CTIO Research Experience for Undergraduates (REU) participant}
\altaffiltext{3}{2004 Gemini South internship, University of Victoria}

\begin{abstract}
We present the calibration of the spectroscopic Lick/IDS standard line-index system for  measurements obtained with the Gemini Multi-Object Spectrographs known as GMOS-North and GMOS-South.~We provide linear correction functions for each of the 25 standard Lick line indices for the B600 grism and two instrumental setups, one with 0.5\arcsec\ slit width and $1\!\times\!1$ CCD pixel binning (corresponding to $\sim\!2.5$\,\AA\ spectral resolution) and the other with 0.75\arcsec\ slit width and $2\!\times\!2$ binning ($\sim\!4$\,\AA). We find small and well-defined correction terms for the set of Balmer indices H$\beta$, H$\gamma_A$, and H$\delta_A$ along with the metallicity sensitive indices Fe5015, Fe5270, Fe5335, Fe5406, Mg$_2$ and Mg$b$ that are widely used for stellar population diagnostics of distant stellar systems.~We find other indices that sample molecular absorption bands, such as TiO$_1$ and TiO$_2$, with very wide wavelength coverage or indices that sample very weak molecular and atomic absorption features, such as Mg$_1$, as well as indices with particularly narrow passband definitions, such as Fe4384, Ca4455, Fe4531, Ca4227, and Fe5782, less robustly calibrated. These indices should be used with caution.
\end{abstract}

\keywords{Methods: data analysis, Techniques: spectroscopic}

\section{Introduction}
The spectroscopic Lick index system was introduced by \cite{burstein84} in order to homogenize the study of low-resolution integrated-light spectra of elliptical galaxies and other extragalactic stellar systems.~It is based on observations carried out with the Cassegrain image dissector scanner (IDS) spectrograph at the 3-m Shane telescope of the Lick Observatory and has been continuously updated and refined by several subsequent works \citep[e.g.][]{worthey94etal, worthey97, trager98}.~In its current form it defines 25 widely-used line indices for specific atomic and molecular absorption features in the optical wavelength range from $\sim\!4040$ to $\sim\!6420$~\AA.~The precise definition of line index passbands allows a reproducible measurement and uniform interpretation of spectroscopic data, in particular of index combinations sensitive to luminosity-weighted stellar population age and various chemical abundances. Because the Lick system was initially devised to study the integrated-light spectra of massive early-type galaxies with high velocity dispersions \citep{burstein84, worthey94, worthey97}, the spectral indices are defined for low spectral resolution at about $R\!\la\!700$ which is equivalent to a spectral resolution of $8\!-\!11$ \AA\ in the optical, \citep[see Fig.~7 in][]{worthey97}.~However, higher resolution index definitions which are formally more sensitive to the absorption feature of interest \citep[e.g.][]{vazdekis10} deliver lower signal-to-noise with the same instrumental setup and exposure time.~In general, spectral resolution and integration times have to be traded for any index system to yield the most efficient observing program and most robust index measurements depending on the target luminosity.

That said, the great advantage of the Lick system is its comprehensive library of nearby-star stellar spectra covering a large parameter space in $\log g, T_{\rm eff}$,~and metallicity.~This library is the foundation of many population synthesis models that use fitting functions computed from this library \citep{tripicco95} to model predictions of Lick index strengths as a function stellar-population age and chemical composition for simple and composite stellar populations \citep[see e.g.][]{trager00, thomas03, thomas04}.~One can then use such model predictions to derive ages and chemical makeups of distant stellar populations, such as galaxies \citep[e.g.][]{thomas05} and globular clusters \citep[e.g.][]{puzia05}, provided the observed spectra match the spectroscopic characteristics of the IDS Lick spectrograph in terms of wavelength coverage, spectroscopic resolution, and continuum flux calibration. Calibrating spectroscopic index measurements onto the Lick system is usually done by re-observing stars from the Lick standard star library \citep{worthey94etal} with the same instrumental configuration that is being used for science observations \citep[see e.g.][]{puzia02}.~The purpose of this work is to provide such Lick index calibrations for the Gemini Multi-Object Spectrographs at the Gemini North and Gemini South Observatory for the most commonly used B600 grating and spectrograph setups.

\section{Observations}
\label{ln:obs}
All spectroscopic observations were collected using both telescopes of the Gemini Observatory.~The first campaign was conducted with the Gemini Multi-Object Spectrograph (GMOS) on the 8.2-m Gemini-North telescope on Mauna Kea in Hawaii (hereafter GMOS-N) as part of the programs GN-2002A-Q-17, GN-2002B-Q-77, GN-2003A-Q-75, and GN-2004A-Q-94.~The second campaign took place at the 8.2-m Gemini-South telescope on Cerro Pach\'{o}n in Chile (hereafter GMOS-S) as part of programs GS-2003B-Q-63 and GS-2004A-Q-62 (see Table~\ref{tab:Nobs} for a log of all GMOS observations).

\begin{deluxetable*}{lccccccc}[!ht]
\tabletypesize{\scriptsize}
\tablecaption{Journal of GMOS-N and GMOS-S Observations \label{tab:Nobs}}
\tablewidth{0pt}
\tablehead{
\colhead{Star} &  \colhead{Program} & \colhead{Observing Date} & \colhead{Slit width} & 
\colhead{Binning} & \colhead{Grating} & \colhead{Blaze $\lambda$ [nm]} & \colhead{$t_{\rm exp}$ [sec]}
}
\startdata
HD074377& GN-2002A-Q-17 & 2002-02-12 &  0.75\arcsec\ &$2\times2$ & B600\_G5303& 508,510     & 120,60 \\ %1
HD148816& GN-2002A-Q-17 & 2002-02-15 &  0.75\arcsec\ &$2\times2$ & B600\_G5303& 508,510     & 5,5 \\ %1
HD165195& GN-2002A-Q-17 & 2002-03-09 &  0.75\arcsec\ &$2\times2$ & B600\_G5303& 508,510     & 20,20 \\ %1
HD172401& GN-2002A-Q-17 & 2002-03-09 &  0.75\arcsec\ &$2\times2$ & B600\_G5303& 508,510     & 20,20 \\ %3
HD172958& GN-2002A-Q-17 & 2002-03-14 &  0.75\arcsec\ &$2\times2$ & B600\_G5303& 508,510     & 20,20 \\ %1
HD199580& GN-2002A-Q-17 & 2002-04-17 &  0.75\arcsec\ &$2\times2$ & B600\_G5303& 508,510,510 & 10,10,5 \\ %1
\\\hline\\
HD224930& GN-2002B-Q-77 & 2002-10-08 &  0.5\arcsec\ &$1\times1$ & B600\_G5303& 2x(508,512) & 4x30 \\ %2
HD017709& GN-2002B-Q-77 & 2002-10-08 &  0.5\arcsec\ &$1\times1$ & B600\_G5303& 508,512     & 45,45 \\ %2
HD034411& GN-2002B-Q-77 & 2002-10-08 &  0.5\arcsec\ &$1\times1$ & B600\_G5303& 508         & 4 \\  %2
HD224930& GN-2002B-Q-77 & 2002-10-09 &  0.5\arcsec\ &$1\times1$ & B600\_G5303& 508,512,4x508  & 30,30,120,3x300 \\ %2
HD019373& GN-2002B-Q-77 & 2002-10-09 &  0.5\arcsec\ &$1\times1$ & B600\_G5303& 508,512,512 & 120,60,120 \\ %1
HD020893& GN-2002B-Q-77 & 2002-10-09 &  0.5\arcsec\ &$1\times1$ & B600\_G5303& 508         & 120 \\\vspace{-0.2cm} %2
\\
HD097907& GN-2003A-Q-75 & 2003-04-28 &  0.5\arcsec\ &$1\times1$ & B600\_G5303& 508,512     & 180,300 \\ %1
HD143761& GN-2003A-Q-75 & 2003-04-28 &  0.5\arcsec\ &$1\times1$ & B600\_G5303& 508,512     & 180,180 \\ %1
HD147677& GN-2003A-Q-75 & 2003-04-28 &  0.5\arcsec\ &$1\times1$ & B600\_G5303& 508,512     & 300,300 \\ %1
HD172401& GN-2003A-Q-75 & 2003-06-02 &  0.5\arcsec\ &$1\times1$ & B600\_G5303& 508,512     & 120,120 \\ \vspace{-0.2cm} %3
\\      
HD161817& GN-2004A-Q-94 & 2004-06-15 &  0.5\arcsec\ &$1\times1$ & B600\_G5303& 2x(508,512) & 4x60   \\ %1
HD168720& GN-2004A-Q-94 & 2004-06-15 &  0.5\arcsec\ &$1\times1$ & B600\_G5303& 508,512     & 60,60   \\ %1
\\\hline\\
HD200779& GS-2003B-Q-63&  2003-09-27&     0.5\arcsec&     $1\times1$&     B600\_G5323&    508,512&        60.5,60.5 \\ %1
HD207076& GS-2003B-Q-63&  2003-09-27&     0.5\arcsec&     $1\times1$&     B600\_G5323&    508,512&        90.5,90.5 \\ %1
HD219617& GS-2003B-Q-63&  2003-09-27&     0.5\arcsec&     $1\times1$&     B600\_G5323&    508,512&        90.5,90.5 \\ %2
HD175638& GS-2003B-Q-63&  2003-10-01&     0.5\arcsec&     $1\times1$&     B600\_G5323&    508,512&        60.5,60.5 \\ %1
HD184406& GS-2003B-Q-63&  2003-10-01&     0.5\arcsec&     $1\times1$&     B600\_G5323&    508,512&        60.5,60.5 \\ %64                                      
HD184492& GS-2003B-Q-63&  2003-10-01&     0.5\arcsec&     $1\times1$&     B600\_G5323&    508,512&        60.5,60.5 \\ %3                                   
HD036003& GS-2003B-Q-63&  2003-10-02&     0.5\arcsec&     $1\times1$&     B600\_G5323&    508,512&        60.5,60.5 \\ %1
HD064606& GS-2003B-Q-63&  2003-11-01&     0.5\arcsec&     $1\times1$&     B600\_G5323&    508,512&        60.5,60.5 \\ %1                                    
HD043318& GS-2003B-Q-63&  2003-11-02&     0.5\arcsec&     $1\times1$&     B600\_G5323&    508,512&        60.5,60.5 \\ %2                                       
HD037160& GS-2003B-Q-63&  2003-11-02&     0.5\arcsec&     $1\times1$&     B600\_G5323&    508,512&        60.5,60.5 \\ %2                                    
HD049161& GS-2003B-Q-63&  2003-11-02&     0.5\arcsec&     $1\times1$&     B600\_G5323&    508,512&        60.5,60.5 \\ %1                                     
HD069267& GS-2003B-Q-63&  2003-11-27&     0.5\arcsec&     $1\times1$&     B600\_G5323&    508,512&        60.5,60.5 \\ \vspace{-0.2cm} %3
\\
HD145148& GS-2004A-Q-62&  2004-04-21&     0.5\arcsec&     $1\times1$&     B600\_G5323&    508,512&        60.5,60.5 \\ %2
HD165760& GS-2004A-Q-62&  2004-07-20&     0.5\arcsec&     $1\times1$&     B600\_G5323&    508,512&        120.5,120.5\\ %40
\enddata
\tablecomments{For the GMOS-N observations setting A ($2\!\times\!2$ binning and 0.75\arcsec\ wide slits) we obtained observations for 6 stars with 13 individual exposures, while for setting B ($1\!\times\!1$ binning and 0.5\arcsec\ slit width) we observed 11 stars with 35 individual exposures.~For the GMOS-S observations in setting B ($1\!\times\!1$ binning and 0.5\arcsec\ slit width) we observed 10 stars with 20 individual exposures.}
\end{deluxetable*}

\subsection{Sample Selection}
The stars observed by us are so-called 'secondary Lick standards' that were used to derive the fitting functions in many population synthesis models \citep{worthey94, faber85, trager98}.~Since most of these sample stars were observed only 1-3 times at Lick observatory (except for HD184406 and HD165760 which were observed 64 and 40 times, respectively) part of the scatter in the calibrations will be associated with the uncertainties in the Lick measurements themselves.~Our star sample is a subset of the total Lick/IDS standard star list selected by H.~Kuntschner for the SAURON integral field spectrograph project \citep{kuntschner06}.~The selection criteria are: 1) $-10^{\rm o}\!<\! {\rm Dec} \!<\! 70^{\rm o}$; 2) the star must be part of the Henry Draper (HD) Catalogue \citep[see][and references therein]{nesterov95}; and 3) the star must not be classified as peculiar. There are 233 stars in this subset spread across all R.A.~coordinates. The SAURON project observed 73 of these stars as part of its survey of the dynamics and stellar populations of nearby early-type galaxies.~Our sub-sample summarized in Table~\ref{tab:Nobs} facilitates therefore a comparison and cross-calibration option between SAURON and GMOS, and previous results from other instruments.

\subsection{Spectrograph Configuration}
At the time of our observations the twin GMOS instruments provided a wavelength coverage within $3600\!-\!9400$ \AA\ in long-slit and multi-slit mode over a $5.5\arcmin\!\times\!5.5\arcmin$ field of view. The instruments are equipped with three EEV CCDs with 2.8\arcsec\ gaps between the detector chips. Since the gaps run perpendicular to the dispersion direction, we took dithered sub-integrations with two different grating blaze angles (at $\lambda\!=\!508$ and 510 nm) to cover the entire wavelength range.~All standard star observations were performed in long-slit mode with the B600 grating with a wavelength coverage from $3800\!-\!6500$ \AA\ with a resolution of $R\approx1500$. The ``central spectrum'' region of interest on the CCD detector was used to place the standard star along the spatial axis of the slit and to reduce the size of the corresponding files.

For the GMOS-N observations we used two instrumental setups with different detector chip binning factors and slit widths.~In the following we refer to setting A as the configuration with $2\!\times\!2$ detector binning and 0.75\arcsec\ slit width (6 stars with 13 individual observations) and to setting B with $1\times1$ binning and 0.5\arcsec\ slit width (11 stars with 35 individual observations).~Most of the GMOS-N observations were conducted using the standard procedure of centering the target in the slit, however for GN-2003A-Q-75 the observations were unguided and the telescope was nodded such that the light from the stars moved across the narrow dimension of the slit aperture (perpendicular to the spatial axis and parallel to the dispersion axis).~With this technique the light from the star fills the slit and the spectral resolution depends only on the slit width.~The detectors were the original E2V devices EEV9273-16-03, EEV9273-20-04, and EEV9273-20-03.~All observations at GMOS-S were conducted with the B setting (14 stars with 28 individual exposures).~The detectors were E2V devices EEV 2037-06-03, EEV 8194-19-04, and EEV 8261-07-04.~All the GMOS-S observations were carried out with the nodding method described above.

\section{Data Reduction}
\label{ln:dr}
The data were reduced using common procedures for long-slit data.~The Gemini GMOS IRAF package was used for most reduction steps.~These scripts are mostly wrappers for standard IRAF\footnote{IRAF is distributed by the National Optical Astronomy Observatories, which are operated by the Association of Universities for Research in Astronomy, Inc., under cooperative agreement with the National Science Foundation.} tasks that handle the Gemini Multi-Extension FITS (MEF) data format.~Differences from standard Gemini procedures are noted below.

\subsection{Bias Subtraction and Flat Fielding}
Bias subtraction was performed using average bias frames (0 second exposure time) taken the closest in time to a given Lick dataset.~It proved best to subtract the overscan from the average biases and the science data as the bias level drifts slowly with time.~Care must be taken to use only the area of overscan that is farthest from the detector area so that the overscan is free from any contamination.~This helps reducing artificial "jumps" in the spectra due to bias subtraction.

Flat field frames were created from quartz halogen lamp spectra taken before or after the Lick star exposures.~The quartz lamp signature was removed using a wrapper script for the IRAF {\sc response} task.~Since the effective slit length was shortened by using the "central spectrum" region of interest, twilight flats were not used to apply an illumination correction.

\subsection{Wavelength Calibration}
Wavelength calibration was done by fitting $4^{\rm th}$-order Chebyshev polynomials to the positions of lines identified in CuAr arc lamp spectra, taken close in time of the stellar observations. The rms of the wavelength solutions was typically 0.2\,\AA. The wavelength solutions were measured on the 2D spectra using the IRAF tasks {\sc autoidentify} and {\sc reidentify} and applied to the Lick standard star observations using the task {\sc transform}.

\subsection{Scattered Light Subtraction}
GMOS data suffer from scattered light, especially for central wavelengths shorter than 5200 \AA. After the spectra were wavelength calibrated the IRAF task {\sc apscatter} was used to fit the diffuse background light in each frame. Particular care was taken to fit and subtract the background light and not the light from the wings of the PSF.~Fit orders of 11 in X and 7 in Y gave good results. This step significantly improved the results for several of the stars, especially those of earlier spectral types.~We point out that since the scattered-light component is a smoothly varying function of detector position that changes on spatial scales much larger than the individual absorption features measured by a Lick index, the contribution to the total Lick index uncertainty associated with this subtraction is negligible compared to variations in the sky subtraction.

\subsection{Quantum Efficiency Corrections}
The three CCDs in each GMOS detector array were chosen to have similar quantum efficiency (QE) characteristics but there are still differences of a few percent that depend on wavelength.~A method was developed to measure the relative QE curves of CCDs 1 and 3 relative to CCD 2.~A correction to the QE differences as a function of wavelength can then be applied.~This can further reduce or remove intensity jumps at the chip boundaries.~This correction step is now being implemented in the Gemini GMOS IRAF package.

\subsection{Parallactic Angle Slit-Loss Corrections}
The GMOS instruments do not have atmospheric dispersion correctors (ADCs) so unless a slit is placed along the parallactic angle at the time of the observation there will be wavelength-dependent slit losses \citep{filippenko82}.~An IDL procedure was developed to calculate the slit losses based on the difference between the position angle (PA) of an observation and the parallactic angle and a correction applied to the spectra.~While this does not restore the lost flux it is important for making the continuum shapes from different observations similar enough that these spectra can be successfully averaged.

\subsection{Flux Calibration}
Relative flux calibrations were performed with observations of flux standard stars and analyzed with the IRAF tasks {\sc standard} and {\sc sensfunc} (via the Gemini task {\sc gsstandard}).~The GMOS-N data were calibrated using observations of EG131 taken in June 2002.~The mean atmospheric extinction curve for Mauna Kea from the Gemini IRAF package was used.~The GMOS-S data were calibrated using an observation of LTT~9239 from November 2003.~The atmospheric extinction curve for Cerro Paranal obtained from the ESO web site was used for the GMOS-S data.

Each flux-calibrated spectrum was compared by eye with reference spectra of similar spectral type from the MILES library \citep{vazdekis10} and duplicate observations of the same stars obtained with VLT/FORS \citep{puzia05}.~We define the final sample of high-quality GMOS Lick standard-star spectra as those spectra for which we find flux discrepancies smaller than 20\% between their final flux calibration and the reference spectra mentioned above.~The final sample spectra can be downloaded from the GMOS calibration web page \href{http://www.gemini.edu/node/10697}{http://www.gemini.edu/node/10697}.

\subsection{Radial Velocity Measurements}
\label{ln:rv}
To derive the rest-frame radial velocity of each Lick standard star we use the spectroscopic stellar library, STELIB\footnote{Spectra are available at http://webast.ast.obs-mip.fr/stelib.} which consists of a homogeneous set of stellar spectra with various spectral types, luminosity classes, and metallicities in the wavelength range 3200-9500 \AA, with a spectral resolution of $\sim\!3$ \AA\ \citep[see][]{lebrogne03}.~The overall absolute photometric uncertainty of this library is 3\% and it provides an excellent reference frame in terms of spectral type and spectral resolution, matching the properties of our GMOS spectra and allowing us to derive accurate radial velocity measurements while testing the influence of spectral mismatch.

We use the {\sc fxcor} package in the IRAF environment \citep{tody93} to derive radial velocities over the full wavelength coverage of our Lick standard spectra. We filter both object and template star spectra with a Welch filter\footnote{After extensive experimentation with the available Fourier filtering functions within {\sc fxcor} we converged on the Welch filter with the following parameters: {\sc cuton}=20, {\sc cutoff}=1000, {\sc fullon}=30, and {\sc fulloff}=800.} and fit the cross-correlation function peak with a Gaussian.~Recorded radial velocities are the corresponding values at the peak of the cross-correlation function.

Mismatch in spectral type between object and template spectrum translates in increased radial velocity errors and we determine the impact of this systematic uncertainty in the following analysis.~For the full sample of Lick standard star spectra we compute the radial velocity mean and dispersion for each star using the entire STELIB library and reject clear outliers in radial velocity space using Chauvenet criteria\footnote{Based on the mean and standard deviation of $n$ data points, a measurement can be discarded as an outlier if its normal distribution probability is less than $(2n)^{-1}$ \citep[see e.g.][]{peirce52}.}. The typical dispersion before the rejection of outliers is then $10\!-\!20$ km/s which is mainly driven by the contribution of template star mismatch. The total error budget is $\sigma_{\rm total}^2 = \sigma_{\rm stat.}^2\!+\! \sigma_{\rm templ.}^2$ and is a linear combination of statistical and template mismatch error. Reducing the sample of template stars to only those within two spectral types of the Lick standard spectral type, we find that the dispersion decreases significantly below 10 km/s.~The resulting error of the mean radial velocity is in all cases $\la\!1$ km/s.~We conclude that selecting template stars with similar spectral types as the Lick standard stars is critical for radial velocity measurements with the cross-correlation technique and for a robust interpretation of the results at the instrumental resolutions probed by this study (see Sect.~\ref{ln:obs}).

\section{Analysis}

\subsection{Spectral Resolution}
\label{ln:specres}

\begin{figure}[!ht]
  \centering
   \includegraphics[width=8cm, bb=0 10 400 385]{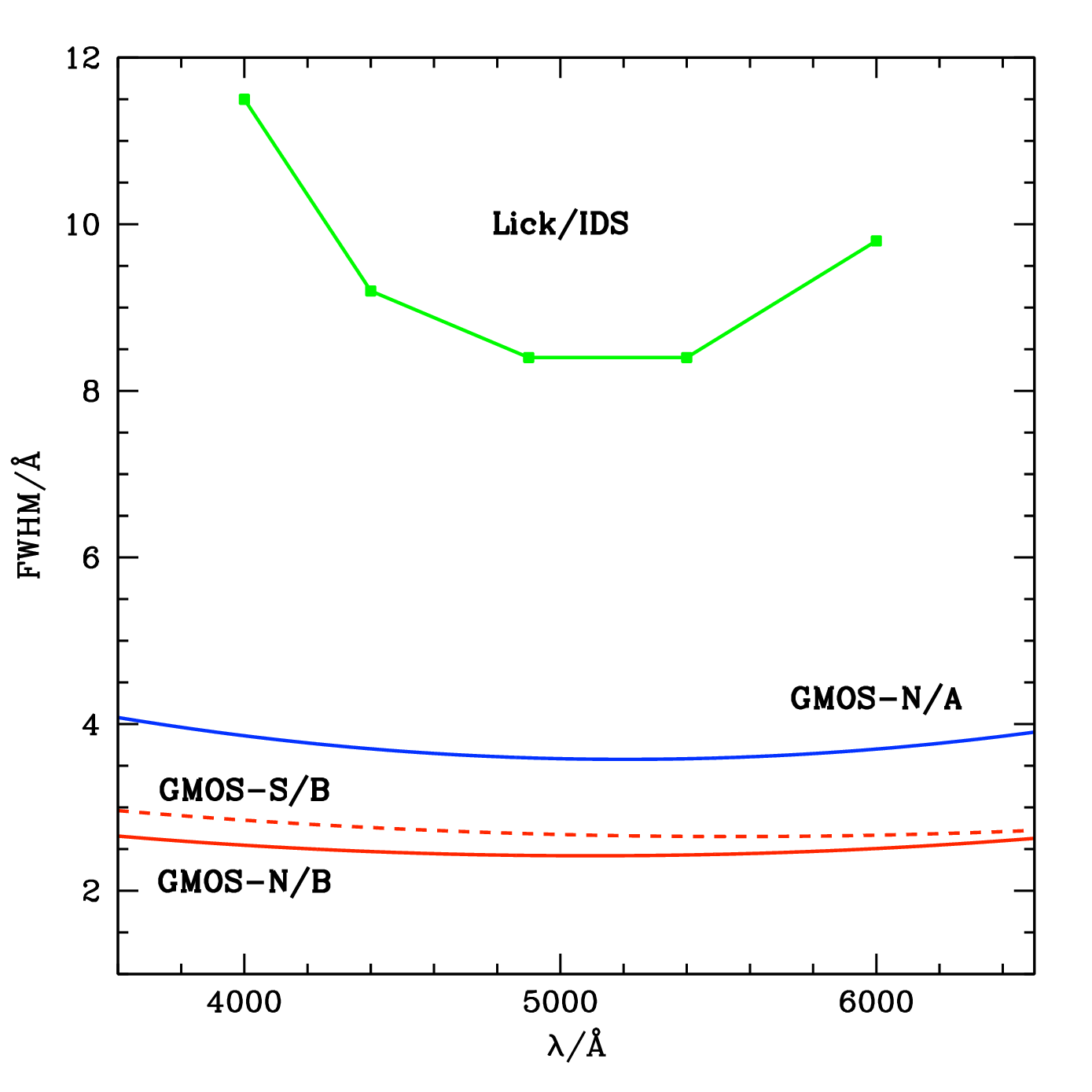} 
\caption{Comparison of the Lick/IDS spectral resolution which is implicitly hardwired in the Lick index system and the corresponding resolutions for our three GMOS instrument configurations.}
   \label{fig:specresol}
\end{figure}

The spectral resolution of our GMOS-N observations for the two instrument configurations was determined using the FWHM of wavelength-calibration lamp emission lines and found to be very well approximated by polynomial functions for the instrument setup A
($2\!\times\!2$--0.75\arcsec) and setup B ($1\!\times\!1$--0.5\arcsec):
\begin{align*}
{\rm FWHM}_{\rm i} (\lambda)=a+b\lambda + c\lambda^2
\end{align*}
where FWHM$_i(\lambda)$ is for the $i$-th setup. $\lambda$ is in units of \AA. Table~\ref{tab:ffresol} summarizes the coefficients and quantifies the fit quality for all instrumental setups.~Note that for instrument setup B the spectral resolutions of GMOS-N and GMOS-S are relatively similar and that all three GMOS configurations are significantly smaller than the nominal spectral resolution of the Lick index systems \citep{worthey97}.~The spectral resolutions of the various GMOS configurations and the Lick/IDS resolution are shown in Figure~\ref{fig:specresol}.~These analytic functions serve as baseline to broaden the GMOS spectra to the Lick/IDS spectral resolution before applying the index measuring routines.~The smoothing of our sample spectra is performed with a wavelength-dependent Gaussian kernel with
\begin{equation*}
\sigma_{\rm smooth}(\lambda)=\left(\frac{\mathrm{FWHM}(\lambda)_{\rm Lick}^2- \mathrm{FWHM}(\lambda)_{\rm i}^2}{8\, {\rm ln}2}\right)^{\frac{1}{2}}
\end{equation*}
where FWHM$_{\rm i}$ describes the spectral resolution of the corresponding instrument setup.

\begin{deluxetable}{lcccc}[!ht]
\tabletypesize{\scriptsize}
\tablecaption{Fitting functions of GMOS spectral resolution \label{tab:ffresol}}
\tablewidth{0pt}
\tablehead{
\colhead{setup} & \colhead{$a$} & \colhead{$b\cdot10^{-3}$} & \colhead{$c\cdot10^{-7}$} & \colhead{rms} 
}
\startdata
GMOS-N/A  & $8.86\pm0.68$ &  $-2.03\pm0.27$ &  $1.95\pm0.25$ & 0.071 \\
GMOS-N/B  & $5.17\pm0.44$ &  $-1.08\pm0.17$ &  $1.06\pm0.16$ & 0.053 \\
GMOS-S/B  & $5.17\pm0.75$ &  $-0.91\pm0.30$ &  $0.82\pm0.28$ & 0.089 \\\vspace{-0.2cm}
\enddata
\tablecomments{The fitting functions give the spectral resolution in FWHM as a function of wavelength, both in units of \AA.}
\end{deluxetable}

\begin{figure*}[!ht]
  \centering
   \includegraphics[width=8.3cm, bb=68 176 574 687]{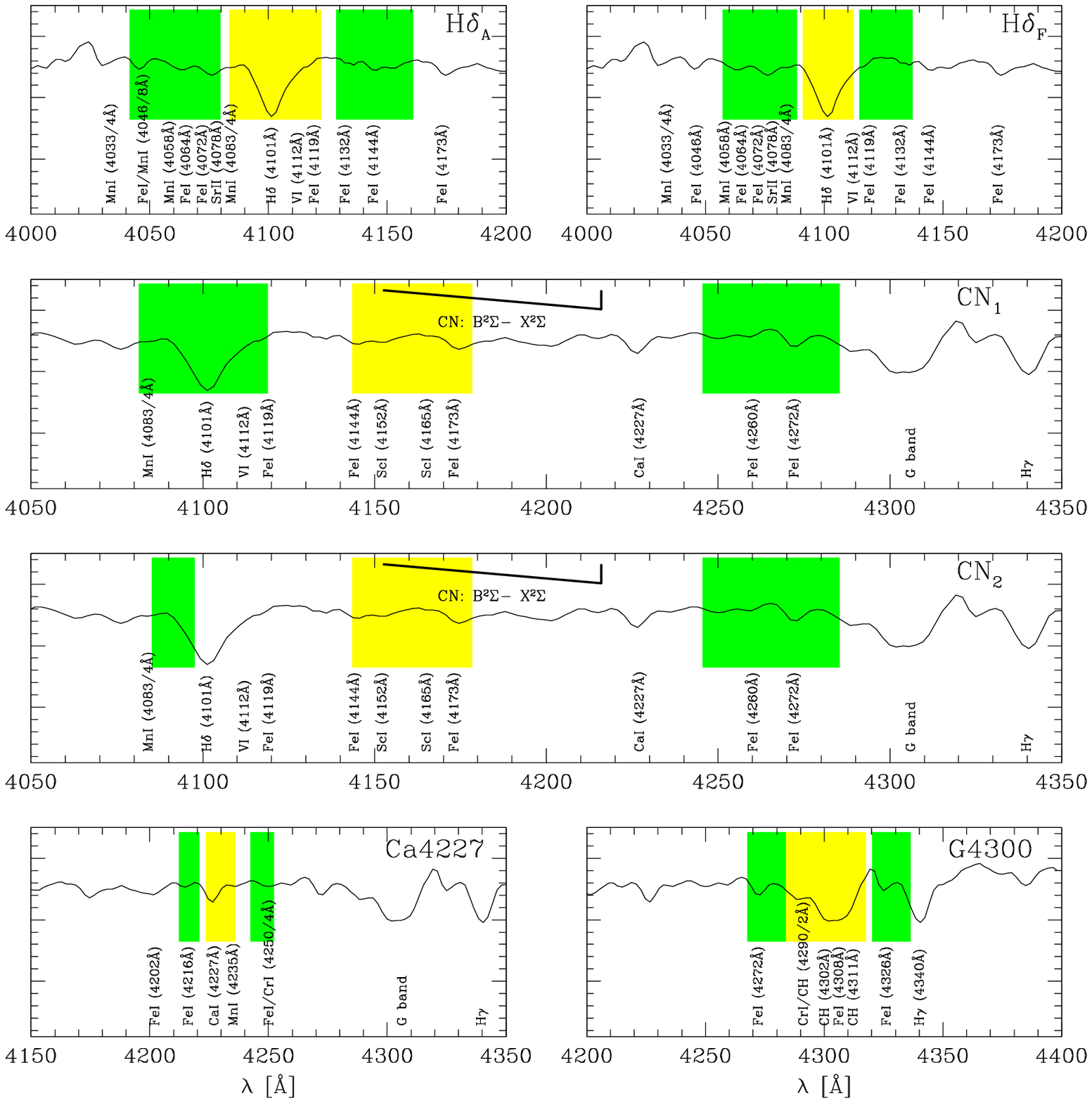} 
   \includegraphics[width=8.3cm, bb=68 176 574 687]{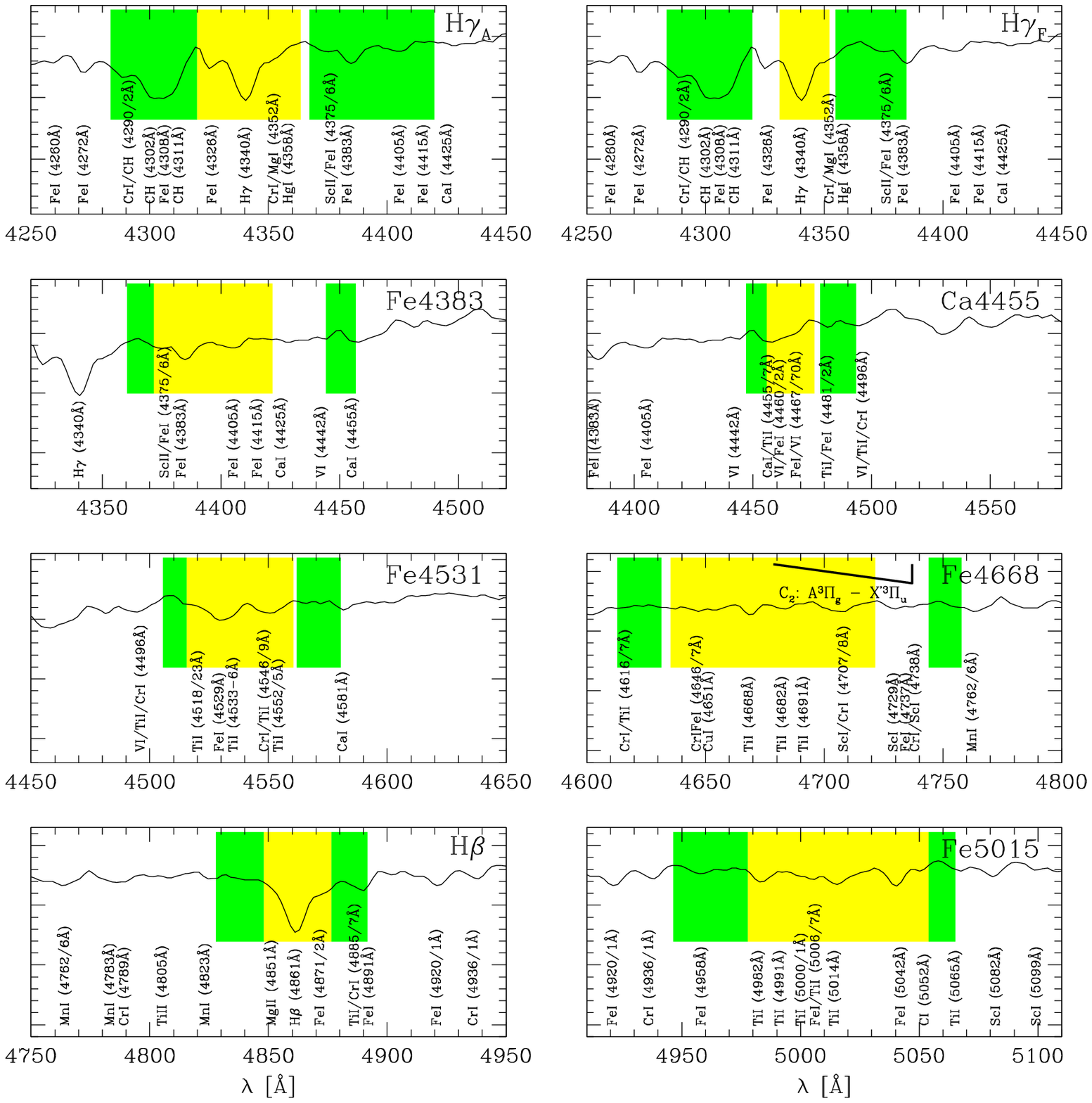} 
   \includegraphics[width=8.3cm, bb=68 176 574 687]{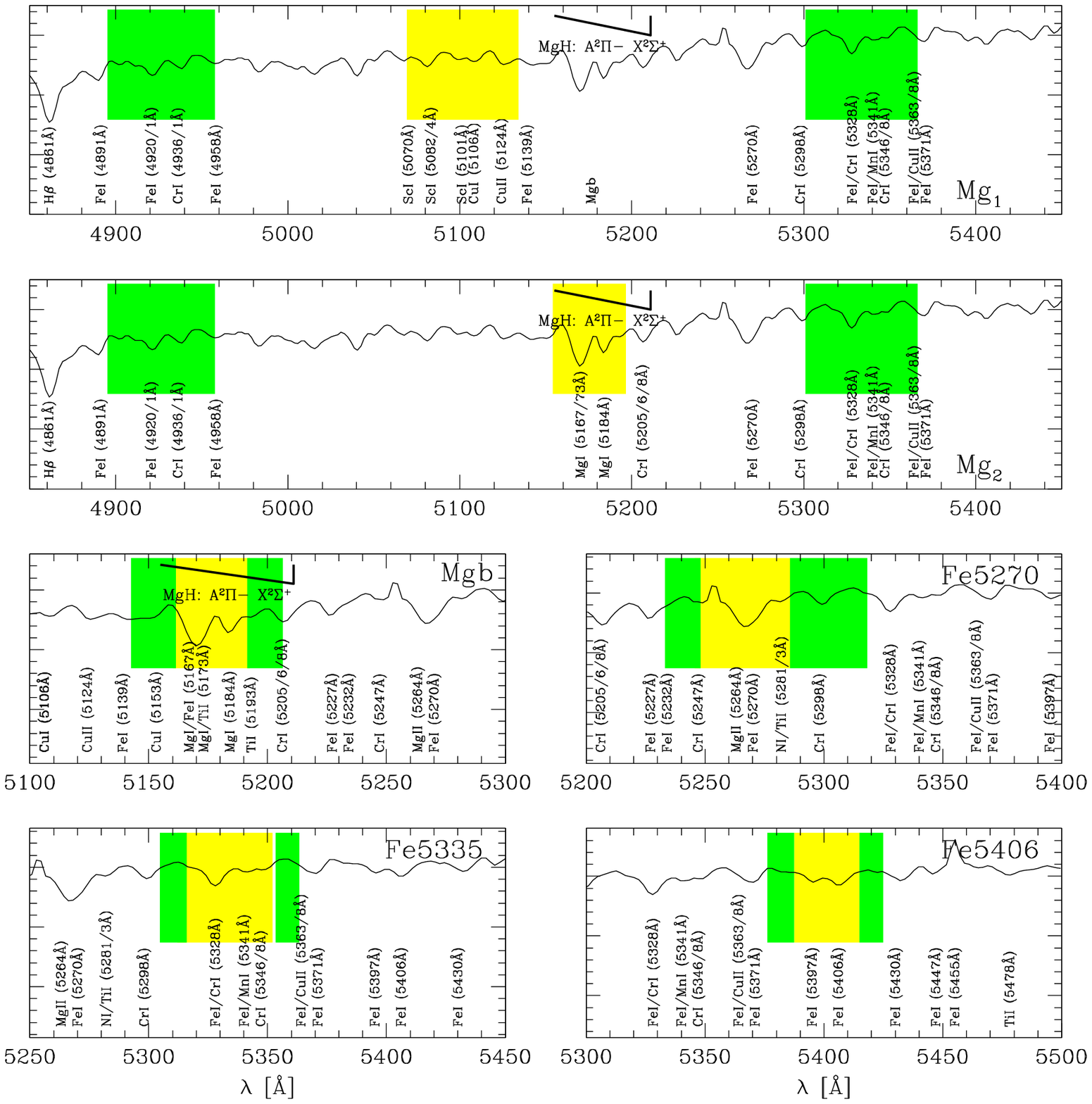} 
   \includegraphics[width=8.3cm, bb=68 176 574 687]{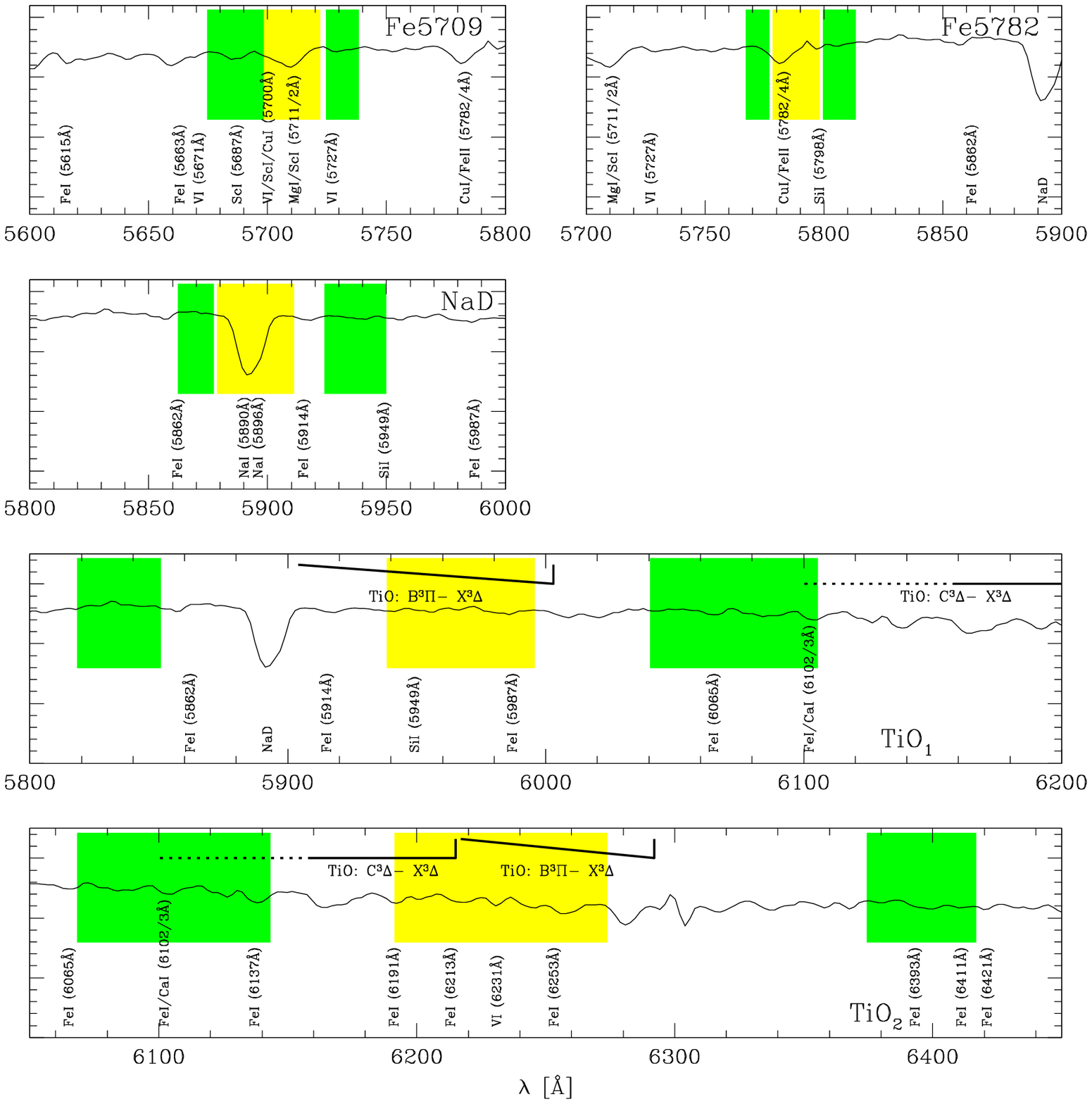} 
\caption{Illustration of Lick index passband definitions for the 25
indices as defined in Table~\ref{tab:pb}. Green and yellow shaded regions
mark the continuum and feature passbands, respectively. Satellite absorption lines as
well as molecular bands are indicated \citep{reader81,pearse76} together
with the spectrum of the Milky Way globular cluster
NGC~6284, taken from \cite{puzia02}.}
   \label{fig:idxpb}
\end{figure*}

\subsection{Lick Index Measurements}
\label{ln:lickidx}

A crucial part of measuring the strength of spectroscopic absorption or emission features is the accurate definition of the adjacent continuum.~The Lick system bypasses this complication by defining a {\it pseudo}-continuum around strong absorption features of interest, as illustrated in Figure~\ref{fig:idxpb}.~To determine the continuum level inside the feature passband a linear interpolation of the mean fluxes in two satellite passbands is performed which defines the {\it pseudo}-continuum flux $F_{\rm c}$.~The feature and {\it pseudo}-continuum passband definitions are summarized in Table~\ref{tab:pb}.

The flux ratio in the feature passband between the absorption line $F_{\rm l}$ and the {\it pseudo}-continuum $F_{\rm c}$ is then used to define the line index
\begin{equation}
\label{eqn:aidxdef}
I_{\rm a}=\int\limits_{\lambda_{\rm min}}^{\lambda_{\rm max}}
\left(1-\frac{F_{\rm l}(\lambda)}{F_{\rm
c}(\lambda)}\right)\,d\lambda
\end{equation}
in units of \AA, where $\lambda_{\rm min}$ and $\lambda_{\rm max}$ define the blue and red boundaries of the feature passband, respectively (see Figure~\ref{fig:idxpb}). Note that if $F_{\rm c}$ were the {\it true} continuum, $I_{\rm a}$ would closely resemble an equivalent width. The subscript ``a'' indicates that the line index definition in Equation~\ref{eqn:aidxdef} is used for narrow atomic absorption features which are calculated in \AA ngstr\o m. For molecular absorption features the line index is defined as
\begin{equation}
\label{eqn:midxdef}
I_{\rm
m}=-2.5\,\log\left[\frac{1}{\Delta_\lambda}\int\limits_{\lambda_{\rm
min}}^{\lambda_{\rm max}} \frac{F_{\rm l}(\lambda)}{F_{\rm
c}(\lambda)}\,d\lambda\right]
\end{equation}
in units of magnitudes\footnote{One exception is the G4300 index which is traditionally measured in \AA, although the strength of the dominant feature is driven by the CH molecule abundance.}, where $\Delta_\lambda=\lambda_{\rm max} - \lambda_{\rm min}$. The transformation between the \AA ngstr\o m and magnitude scale can be performed with the following two equations
\begin{eqnarray}
\label{eqn:idxtrafo}
I_{\rm a}&=&\Delta_\lambda\left(1-10^{-0.4\, I_{\rm m}}\right)\\ 
I_{\rm m}&=&-2.5\, \log\left(1-\frac{I_{\rm a}}{\Delta_\lambda}\right)
\end{eqnarray}

The Lick line index strengths for our sample stars, as defined by Equations~\ref{eqn:aidxdef} and \ref{eqn:midxdef} together with the passband definitions summarized in Table~\ref{tab:pb}, were measured with the {\sc GONZO} code that is described in detail in \cite{puzia02, puzia05}.~The GMOS science and variance spectra are used to generate the uncertainty of the Lick index measurements via Monte Carlo simulations. The indices are re-measured on the Poisson noise-altered spectra. From the distribution of index values the 1-$\sigma$ standard deviation defines the total Lick index uncertainty. Note that instead of transforming the spectrum to the restframe the code shifts the index passbands to the observed redshift to avoid pixel noise correlation effects for narrowly defined indices.

\subsection{Lick Index-System Calibration}
\label{ln:lickcal}
\begin{figure*}[ht!]
  \centering
   \includegraphics[width=18cm]{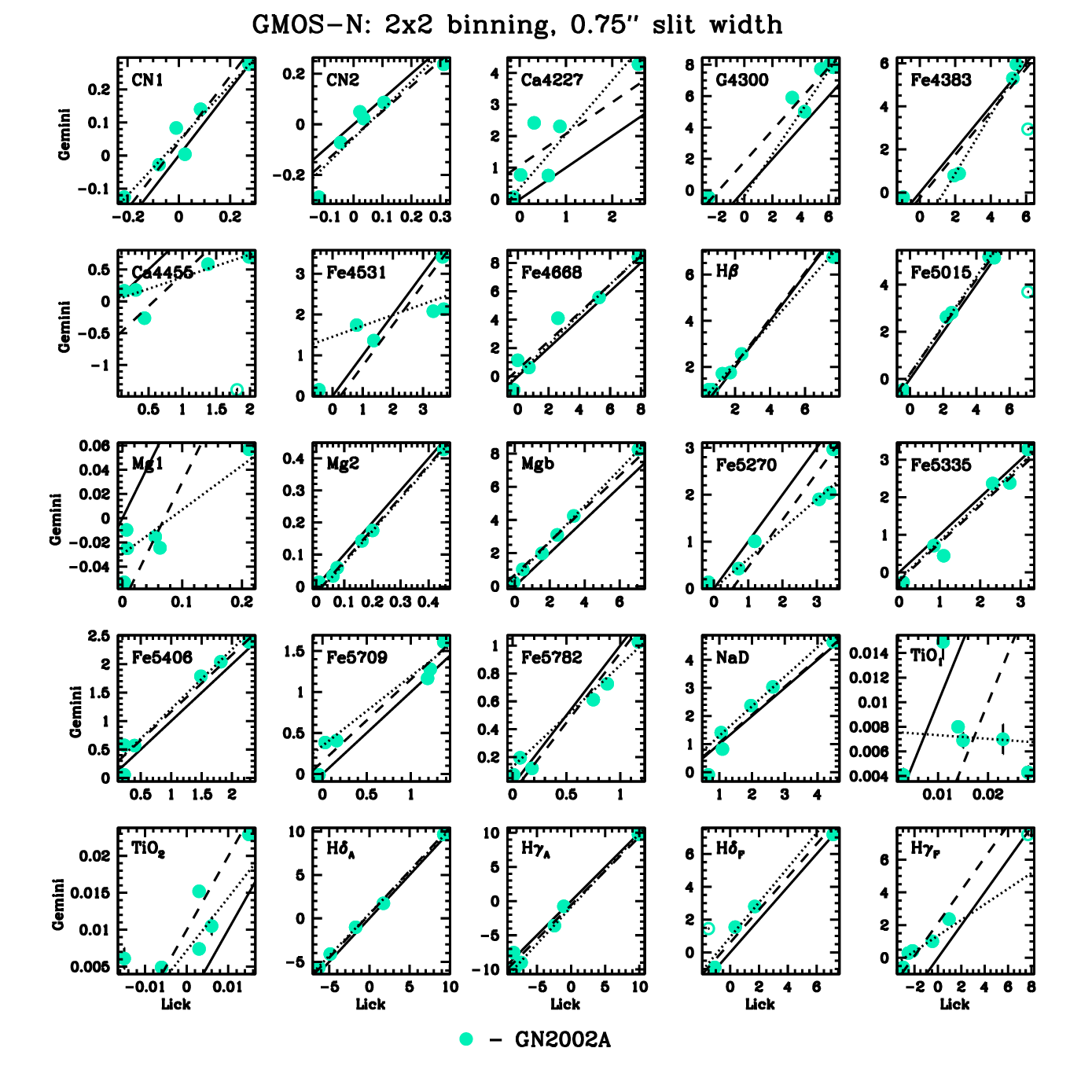} 
   \caption{Comparison plots between original Lick system measurements and Lick index measurements from our standard-star GMOS-N spectra obtained with setup A: $2\!\times\!2$ binning and 0.75\arcsec\ slit width.~In each panel, the solid lines are the identity relations, while dashed lines illustrate the mean linear offset corrections. Dotted lines are weighted least-square fits to the solid data points. Open circles mark data that were rejected using Chauvenet criteria (see text for details).~The corresponding correction terms are summarized in Table~\ref{tab:GN2x2075}. Note that very few data points have error bars larger than the symbol size.}
   \label{fig:compGN2x2}
\end{figure*}

\begin{figure*}[ht!]
  \centering
   \includegraphics[width=18cm]{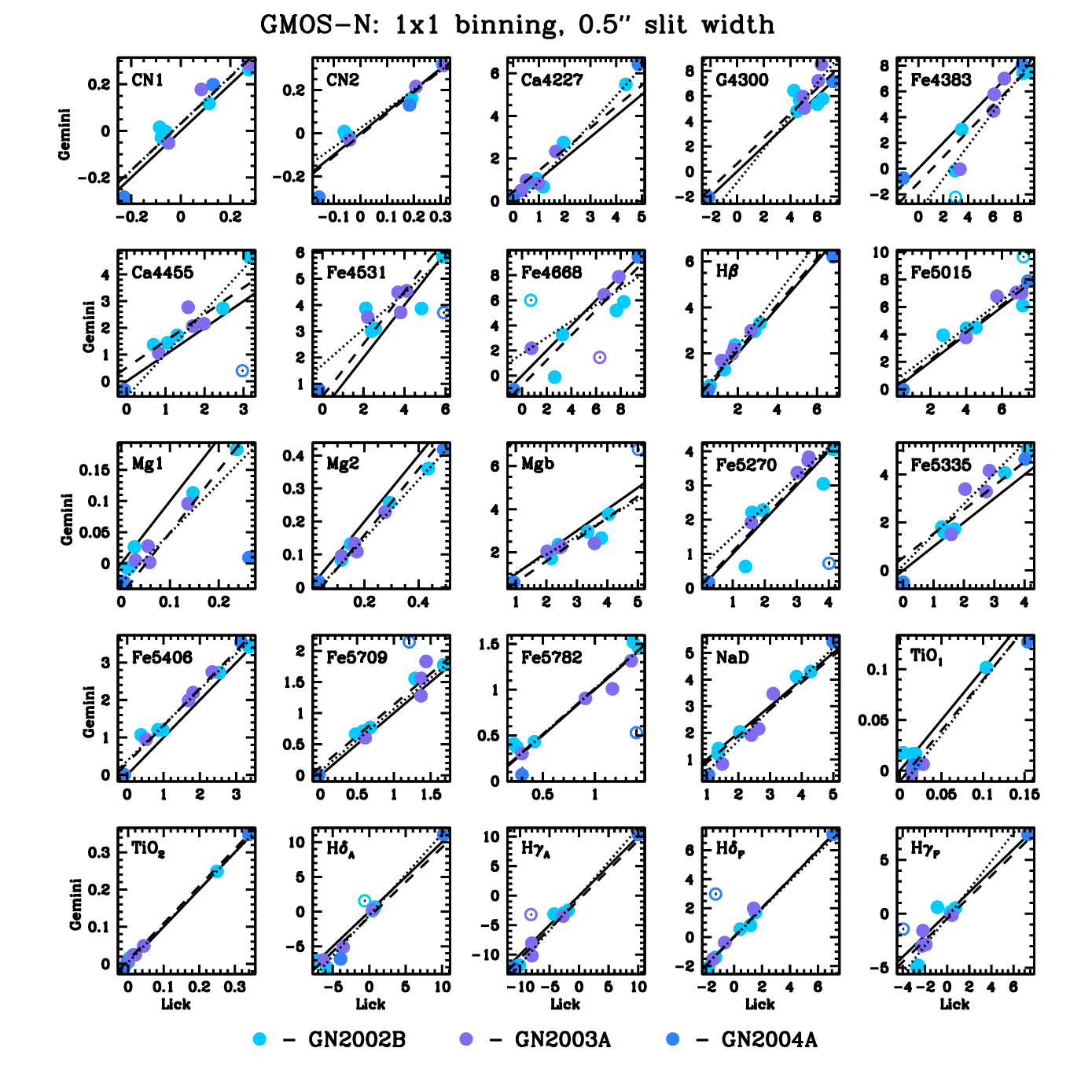} 
   \caption{Comparison plots similar to Figure~\ref{fig:compGN2x2}, but this time for GMOS-N spectra obtained with setup B: $1\times1$ binning and 0.5\arcsec\ slit width.~The different color shadings indicate three different data set from the observing periods 2002B, 2003A, and 2004A. The corresponding correction terms are summarized in Table~\ref{tab:GN1x105}. }
   \label{fig:compGN1x1}
\end{figure*}

\begin{figure*}[ht!]
  \centering
   \includegraphics[width=18cm]{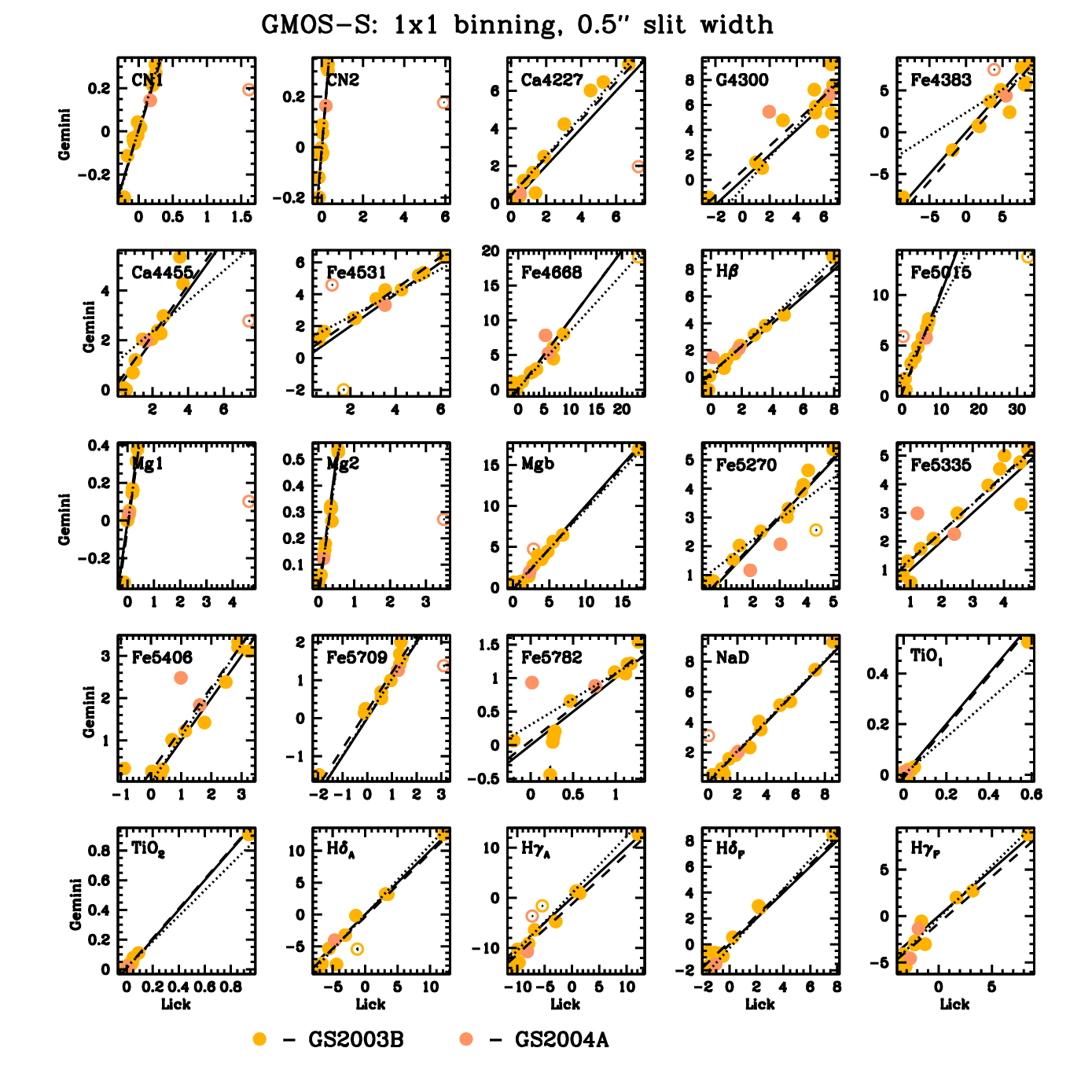} 
   \caption{Comparison plots similar to Figure~\ref{fig:compGN2x2}, but this time for GMOS-S spectra obtained with setup B: $1\times1$ binning and 0.5\arcsec\ slit width.~The different color shadings indicate two different data set from the observing periods 2003B, and 2004A. The corresponding correction terms are summarized in Table~\ref{tab:GS1x105}. }
   \label{fig:compGS1x1}
\end{figure*}

Using the full dataset we determine the mean index correction terms to the Lick system for each instrument setup and calculate the corresponding r.m.s.~For each index, outliers are rejected using Chauvenet criteria as described above.~From the selected data we derive the mean correction in the sense
\begin{equation}
\label{ln:lincorr}
I_{\rm Lick} = I_{\rm GMOS} + \delta
\end{equation}
(see dashed lines in Figures~\ref{fig:compGN2x2}, \ref{fig:compGN1x1} and \ref{fig:compGS1x1}), the error of the mean correction, $\Delta\delta$, and the dispersion, $\sigma(\delta)$. The numerical values of these parameters are summarized in columns two to four of Tables~\ref{tab:GN2x2075}, \ref{tab:GN1x105} and \ref{tab:GS1x105}. We also fit linear relations of the form 
\begin{equation}
\label{ln:eqncorr}
I_{\rm Lick} = a\!\times\! I_{\rm GMOS} + b
\end{equation}
to the selected data (dotted lines in Figures~\ref{fig:compGN2x2}, \ref{fig:compGN1x1} and \ref{fig:compGS1x1}) and determine the uncertainties, $\Delta a$ and $\Delta b$, as well as the root mean square and the range over which the fit is valid.~All parameter values are tabulated in columns five to eleven of Tables~\ref{tab:GN2x2075}, \ref{tab:GN1x105}, and \ref{tab:GS1x105}.~Note that our Lick standard-star sample covers a large dynamic range in each index allowing for a robust correction as a function of index strength.~Figures~\ref{fig:compGN2x2}, \ref{fig:compGN1x1} and \ref{fig:compGS1x1} illustrate the comparison between Lick index system and our measurements for the three instrument configurations. The plots show the overall quality of the Lick index calibration for the three spectroscopic settings.~We point out that although the sampling range of each index varies from epoch to epoch the overall fit quality is not driven by any particular sub-sample. In particular, the most important Lick indices that are typically used in stellar population analyses (marked by bold panels in Figures~\ref{fig:compGN2x2}, \ref{fig:compGN1x1} and \ref{fig:compGS1x1}) show very consistent calibrations.

The previous Lick index calibrations are strictly valid only for stellar systems with zero line of sight velocity dispersion (LOSVD). If objects such as massive galaxies are observed with complex LOSVD distribution functions, higher-order correction terms for the LOSVD broadening of the indices is required. Such correction terms are provided in \cite{kuntschner04}. However, for stellar systems with low LOSVD (i.e. $\sigma<50$ km/s) such as globular clusters and dwarf galaxies these corrections are less than 1\% and can be safely neglected.

\section{Summary} 
\label{summary}
We provide Lick index calibration functions for the GMOS-N and GMOS-S spectrograph using two instrumental configurations, namely with $2\!\times\!2$ binning and 0.75\arcsec\ slit width (setup A) and with $1\!\times\!1$ binning and 0.5\arcsec\ slit width (setup B).~The quality of the linear correction terms shows that widely-used Lick indices such as the Balmer indices H$\beta$, H$\gamma_A$, and H$\delta_A$ along with metallicity sensitive indices Fe5015, Fe5270, Fe5335, Fe5406, Mg$_2$ and Mg$b$ can be robustly calibrated and thus used to derive stellar population parameters by comparison with predictions of stellar population synthesis models that use the exact same Lick index definitions. Indices which sample many weak features or molecular absorption bands, such as Mg$_1$, TiO$_1$ and TiO$_2$, with very wide wavelength coverage or indices with particularly narrow passband definitions, such as Fe4384, Ca4455, Fe4531, Ca4227, and Fe5782, are less robustly calibrated and should be used with caution.

\acknowledgments
We gratefully acknowledge the thoughtful comments and suggestions of the anonymous referee.~T.H.P. acknowledges support by CONICYT through FONDECYT/Regular Project No.~1121005, FONDAP Center for Astrophysics (15010003) and BASAL Center for Astrophysics and Associated Technologies (PFB-06), Conicyt, Chile. Based on observations obtained at the Gemini Observatory, which is operated by the Association of Universities for Research in Astronomy, Inc., under a cooperative agreement with the NSF on behalf of the Gemini partnership: the National Science Foundation (United States), the Science and Technology Facilities Council (United Kingdom), the National Research Council (Canada), CONICYT (Chile), the Australian Research Council (Australia), Minist\'{e}rio da Ci\^{e}ncia, Tecnologia e Inova\c{c}\~{a}o (Brazil) and Ministerio de Ciencia, Tecnolog\'{i}a e Innovaci\'{o}n Productiva (Argentina).

{\it Facilities:} \facility{Gemini (GMOS)}.

\newpage

\appendix
\section{Appendix material}

\begin{deluxetable*}{lcccc}[!ht]
\tabletypesize{\scriptsize}
\tablecaption{Lick index passband definitions \label{tab:pb}}
\tablewidth{0pt}
\tablehead{
\colhead{Index} & \colhead{feature passband} & \colhead{blue continuum} &
\colhead{red continuum} & \colhead{units} }
\startdata
H$\delta_A$ &   4083.500 4122.250 &  4041.600 4079.750 &  4128.500 4161.000 &\AA\\
H$\delta_F$ &   4091.000 4112.250 &  4057.250 4088.500 &  4114.750 4137.250 &\AA\\
CN$_1$      &   4143.375 4178.375 &  4081.375 4118.875 &  4245.375 4285.375 &mag\\
CN$_2$      &   4143.375 4178.375 &  4085.125 4097.625 &  4245.375 4285.375 &mag\\
Ca4227      &   4223.500 4236.000 &  4212.250 4221.000 &  4242.250 4252.250 &\AA\\
G4300       &   4282.625 4317.625 &  4267.625 4283.875 &  4320.125 4336.375 &\AA\\
H$\gamma_A$ &   4319.750 4363.500 &  4283.500 4319.750 &  4367.250 4419.750 &\AA\\
H$\gamma_F$ &   4331.250 4352.250 &  4283.500 4319.750 &  4354.750 4384.750 &\AA\\
Fe4383      &   4370.375 4421.625 &  4360.375 4371.625 &  4444.125 4456.625 &\AA\\
Ca4455      &   4453.375 4475.875 &  4447.125 4455.875 &  4478.375 4493.375 &\AA\\
Fe4531      &   4515.500 4560.500 &  4505.500 4515.500 &  4561.750 4580.500 &\AA\\
Fe4668      &   4635.250 4721.500 &  4612.750 4631.500 &  4744.000 4757.750 &\AA\\
H$\beta$    &   4847.875 4876.625 &  4827.875 4847.875 &  4876.625 4891.625 &\AA\\
Fe5015      &   4977.750 5054.000 &  4946.500 4977.750 &  5054.000 5065.250 &\AA\\
Mg$_1$      &   5069.125 5134.125 &  4895.125 4957.625 &  5301.125 5366.125 &mag\\
Mg$_2$      &   5154.125 5196.625 &  4895.125 4957.625 &  5301.125 5366.125 &mag\\
Mg$b$       &   5160.125 5192.625 &  5142.625 5161.375 &  5191.375 5206.375 &\AA\\
Fe5270      &   5245.650 5285.650 &  5233.150 5248.150 &  5285.650 5318.150 &\AA\\
Fe5335      &   5312.125 5352.125 &  5304.625 5315.875 &  5353.375 5363.375 &\AA\\
Fe5406      &   5387.500 5415.000 &  5376.250 5387.500 &  5415.000 5425.000 &\AA\\
Fe5709      &   5698.375 5722.125 &  5674.625 5698.375 &  5724.625 5738.375 &\AA\\
Fe5782      &   5778.375 5798.375 &  5767.125 5777.125 &  5799.625 5813.375 &\AA\\
NaD         &   5878.625 5911.125 &  5862.375 5877.375 &  5923.875 5949.875 &\AA\\
TiO$_1$     &   5938.375 5995.875 &  5818.375 5850.875 &  6040.375 6105.375 &mag\\
TiO$_2$     &   6191.375 6273.875 &  6068.375 6143.375 &  6374.375 6416.875 &mag\\
\enddata
\tablecomments{The above passband definitions are for the
  full set of 25 Lick indices which are used in this work. The index
  definitions were taken from \cite{worthey94} and \cite{worthey97}. The units
  of each index are given in the last column.}
\end{deluxetable*}

\begin{deluxetable*}{lrrrrrrrrrr}
\tabletypesize{\scriptsize}
\tablecaption{Lick index correction terms for GMOS-N with $2\!\times\!2$ binning and 0.75\arcsec\ slitwidth\label{tab:GN2x2075}}
\tablewidth{0pt}
\tablehead{
\colhead{Index} & \colhead{$\delta$} & \colhead{$\Delta\delta$} & \colhead{$\sigma(\delta)$} &
\colhead{$a$} & \colhead{$\Delta a$} & \colhead{$b$} & \colhead{$\Delta b$} &
\colhead{r.m.s.} & \colhead{min} & \colhead{max}
}
\startdata
 H$\delta_A$&$-$0.50 & 0.14 & 0.31 &   0.9679 & 0.0325 &   0.5751 & 0.1159 & 0.1446 & $-$6.28 & 9.25 \\
 H$\delta_F$&$-$0.62 & 0.27 & 0.60 &   1.0557 & 0.2375 &   0.9616 & 0.2977 & 0.4061 & $-$1.08 & 7.15 \\
 CN$_1$     &$-$0.04 & 0.02 & 0.05 &   0.8327 & 0.1112 &   0.0468 & 0.0166 & 0.0331 & $-$0.21 & 0.27 \\
 CN$_2$     &   0.05 & 0.03 & 0.07 &   1.1163 & 0.2299 &$-$0.0559 & 0.0355 & 0.0643 & $-$0.12 & 0.31 \\
 Ca4227     &$-$1.06 & 0.33 & 0.82 &   1.7038 & 0.2406 &   0.3610 & 0.1383 & 0.2745 & $-$0.13 & 2.57 \\
 G4300      &$-$1.84 & 0.26 & 0.64 &   1.3301 & 0.3060 &$-$0.3287 & 1.5129 & 0.5887 & $-$2.56 & 6.33 \\
 H$\gamma_A$&   0.55 & 0.43 & 1.05 &   1.0773 & 0.0765 &$-$0.8482 & 0.3727 & 0.4720 & $-$8.41 & 9.88 \\
 H$\gamma_F$&$-$2.14 & 0.27 & 0.65 &   0.4789 & 0.1119 &   1.3354 & 0.1561 & 0.1950 & $-$2.97 & 0.95 \\
 Fe4383     &   0.24 & 0.39 & 0.95 &   1.3613 & 0.0986 &$-$1.8227 & 0.3556 & 0.2811 & $-$0.94 & 5.45 \\
 Ca4455     &   0.58 & 0.22 & 0.53 &   0.3581 & 0.1586 &   0.0142 & 0.1275 & 0.1498 &    0.14 & 1.99 \\
 Fe4531     &   0.25 & 0.40 & 0.99 &   0.2659 & 0.1394 &   1.4475 & 0.2532 & 0.3116 & $-$0.45 & 3.69 \\
 Fe4668     &$-$0.46 & 0.33 & 0.80 &   1.0582 & 0.1048 &   0.1047 & 0.3123 & 0.4892 & $-$0.27 & 7.84 \\
 H$\beta$   &$-$0.11 & 0.20 & 0.50 &   0.8531 & 0.0552 &   0.4547 & 0.0744 & 0.0921 &    0.48 & 7.61 \\
 Fe5015     &$-$0.23 & 0.09 & 0.21 &   1.0324 & 0.0421 &   0.2096 & 0.1257 & 0.0696 & $-$0.41 & 5.08 \\
 Mg$_1$     &   0.07 & 0.02 & 0.05 &   0.3658 & 0.1210 &$-$0.0299 & 0.0039 & 0.0071 &    0.00 & 0.21 \\
 Mg$_2$     &   0.02 & 0.00 & 0.01 &   1.0037 & 0.0279 &$-$0.0261 & 0.0027 & 0.0033 &    0.01 & 0.45 \\
 Mg$b$      &$-$0.67 & 0.13 & 0.32 &   1.0912 & 0.0244 &   0.5020 & 0.0439 & 0.0648 & $-$0.06 & 7.08 \\
 Fe5270     &   0.53 & 0.25 & 0.62 &   0.6325 & 0.0589 &$-$0.0035 & 0.1085 & 0.1346 & $-$0.16 & 3.48 \\
 Fe5335     &   0.23 & 0.12 & 0.29 &   1.0152 & 0.1011 &$-$0.1927 & 0.1602 & 0.1621 &    0.10 & 3.18 \\
 Fe5406     &$-$0.16 & 0.08 & 0.19 &   1.0458 & 0.0577 &   0.1624 & 0.0574 & 0.0712 &    0.23 & 2.28 \\
 Fe5709     &$-$0.15 & 0.06 & 0.15 &   0.8457 & 0.0665 &   0.3407 & 0.0473 & 0.0767 & $-$0.04 & 1.37 \\
 Fe5782     &   0.05 & 0.05 & 0.12 &   0.7298 & 0.0504 &   0.1282 & 0.0284 & 0.0439 &    0.01 & 1.17 \\
 NaD        &$-$0.04 & 0.19 & 0.47 &   1.0533 & 0.1386 &   0.2238 & 0.2307 & 0.2060 &    0.67 & 4.48 \\
 TiO$_1$    &   0.01 & 0.00 & 0.01 &$-$0.0298 & 0.1490 &   0.0076 & 0.0023 & 0.0018 &    0.00 & 0.03 \\
 TiO$_2$    &$-$0.01 & 0.00 & 0.01 &   0.7194 & 0.3193 &   0.0072 & 0.0018 & 0.0031 & $-$0.02 & 0.02 \\
\enddata
\tablecomments{See Figure~\ref{fig:compGN2x2} for an illustration of the corresponding Lick index corrections that are being used in Equations~\ref{ln:lincorr} and \ref{ln:eqncorr}.}
\end{deluxetable*}

\begin{deluxetable*}{lrrrrrrrrrr}
\tabletypesize{\scriptsize}
\tablecaption{Lick index correction terms for GMOS-N with $1\!\times\!1$ binning 
and 0.5\arcsec\ slitwidth\label{tab:GN1x105}}
\tablewidth{0pt}
\tablehead{
\colhead{Index} & \colhead{$\delta$} & \colhead{$\Delta\delta$} & \colhead{$\sigma(\delta)$} &
\colhead{$a$} & \colhead{$\Delta a$} & \colhead{$b$} & \colhead{$\Delta b$} &
\colhead{r.m.s.} & \colhead{min} & \colhead{max}
}
\startdata
 H$\delta_A$&   0.79 & 0.37 & 1.17 &   1.1717 & 0.0757 &$-$0.4964 & 0.2786 & 0.5903 & $-$6.90 & 10.23 \\
 H$\delta_F$&$-$0.05 & 0.09 & 0.28 &   0.9631 & 0.0950 &$-$0.0115 & 0.1211 & 0.3197 & $-$1.83 & 7.09 \\
 CN$_1$     &$-$0.03 & 0.02 & 0.05 &   1.0086 & 0.1085 &   0.0311 & 0.0175 & 0.0480 & $-$0.23 & 0.28 \\
 CN$_2$     &   0.01 & 0.02 & 0.06 &   0.8878 & 0.0799 &   0.0227 & 0.0145 & 0.0344 & $-$0.16 & 0.31 \\
 Ca4227     &$-$0.42 & 0.17 & 0.58 &   1.3450 & 0.1078 &$-$0.4221 & 0.1431 & 0.2317 &    0.02 & 4.87 \\
 G4300      &$-$0.60 & 0.30 & 0.99 &   1.3118 & 0.4031 &$-$0.8420 & 2.3819 & 0.6974 & $-$2.15 & 7.21 \\
 H$\gamma_A$&   0.71 & 0.30 & 0.99 &   1.1590 & 0.0861 &$-$0.0725 & 0.6055 & 0.7477 & $-$11.07& 10.07 \\
 H$\gamma_F$&   0.49 & 0.35 & 1.15 &   1.3066 & 0.1587 &$-$0.3130 & 0.2959 & 0.5955 & $-$2.69 & 7.30 \\
 Fe4383     &   1.09 & 0.39 & 1.30 &   1.2981 & 0.1648 &$-$3.3554 & 1.0112 & 0.6174 & $-$1.22 & 8.82 \\
 Ca4455     &$-$0.50 & 0.16 & 0.52 &   1.5966 & 0.1529 &$-$0.5676 & 0.2829 & 0.2143 & $-$0.06 & 3.14 \\
 Fe4531     &$-$0.53 & 0.23 & 0.77 &   0.6841 & 0.0868 &   1.7106 & 0.3664 & 0.2408 & $-$0.17 & 5.89 \\
 Fe4668     &   0.76 & 0.43 & 1.43 &   0.6495 & 0.1602 &   1.7043 & 1.0444 & 0.7967 & $-$0.68 & 9.44 \\
 H$\beta$   &$-$0.14 & 0.10 & 0.32 &   1.1268 & 0.0717 &   0.0715 & 0.1299 & 0.1292 &    0.52 & 6.82 \\
 Fe5015     &$-$0.14 & 0.21 & 0.69 &   0.9854 & 0.1688 &   0.5641 & 0.9325 & 0.5689 &    0.47 & 7.46 \\
 Mg$_1$     &   0.05 & 0.02 & 0.07 &   0.7968 & 0.1320 &$-$0.0307 & 0.0122 & 0.0229 &    0.01 & 0.26 \\
 Mg$_2$     &   0.04 & 0.01 & 0.02 &   0.8887 & 0.0651 &$-$0.0282 & 0.0136 & 0.0164 &    0.03 & 0.49 \\
 Mg$b$      &   0.42 & 0.12 & 0.41 &   0.8506 & 0.1208 &   0.1787 & 0.3965 & 0.2749 &    0.93 & 4.05 \\
 Fe5270     &$-$0.08 & 0.15 & 0.50 &   0.8737 & 0.1124 &   0.6210 & 0.3245 & 0.2457 &    0.24 & 4.14 \\
 Fe5335     &$-$0.52 & 0.17 & 0.57 &   1.2776 & 0.2528 &   0.2193 & 0.5998 & 0.5570 &    0.00 & 4.11 \\
 Fe5406     &$-$0.32 & 0.05 & 0.18 &   0.9559 & 0.0417 &   0.3658 & 0.0758 & 0.0912 & $-$0.08 & 3.35 \\
 Fe5709     &$-$0.13 & 0.04 & 0.14 &   0.9953 & 0.1638 &   0.0651 & 0.1981 & 0.1644 & $-$0.02 & 1.68 \\
 Fe5782     &$-$0.01 & 0.04 & 0.13 &   0.9982 & 0.0549 &   0.0087 & 0.0512 & 0.0578 &    0.22 & 1.42 \\
 NaD        &   0.12 & 0.12 & 0.39 &   1.1631 & 0.1124 &$-$0.6259 & 0.2883 & 0.2969 &    1.05 & 5.00 \\
 TiO$_1$    &   0.01 & 0.00 & 0.01 &   1.0604 & 0.1106 &$-$0.0175 & 0.0029 & 0.0052 &    0.00 & 0.15 \\
 TiO$_2$    &$-$0.01 & 0.00 & 0.00 &   0.9829 & 0.0103 &   0.0051 & 0.0005 & 0.0012 & $-$0.01 & 0.34 \\
\enddata
\tablecomments{See Figure~\ref{fig:compGN1x1} for an illustration of the corresponding Lick index correction terms that are being used in Equations~\ref{ln:lincorr} and \ref{ln:eqncorr}. }
\end{deluxetable*}

\begin{deluxetable*}{lrrrrrrrrrr}
\tabletypesize{\scriptsize}
\tablecaption{Lick index correction terms for GMOS-S with $1\!\times\!1$ binning 
and 0.5\arcsec\ slitwidth\label{tab:GS1x105}}
\tablewidth{0pt}
\tablehead{
\colhead{Index} & \colhead{$\delta$} & \colhead{$\Delta\delta$} & \colhead{$\sigma(\delta)$} &
\colhead{$a$} & \colhead{$\Delta a$} & \colhead{$b$} & \colhead{$\Delta b$} &
\colhead{r.m.s.} & \colhead{min} & \colhead{max}
}
\startdata
 H$\delta_A$&   0.40 & 0.36 & 1.25 &   1.0919 & 0.1355 &$-$0.0086 & 0.5216 & 0.8713 & $-$7.11 & 12.11 \\
 H$\delta_F$&$-$0.19 & 0.15 & 0.52 &   1.1447 & 0.0600 &$-$0.2633 & 0.0711 & 0.2086 & $-$1.57 & 7.71 \\
 CN$_1$     &$-$0.01 & 0.01 & 0.04 &   1.0735 & 0.0797 &   0.0121 & 0.0129 & 0.0413 & $-$0.21 & 0.25 \\
 CN$_2$     &$-$0.01 & 0.01 & 0.05 &   1.0736 & 0.0920 &$-$0.0058 & 0.0165 & 0.0459 & $-$0.14 & 0.30 \\
 Ca4227     &$-$0.46 & 0.18 & 0.64 &   1.0411 & 0.0201 &   0.4598 & 0.0447 & 0.1135 &    0.15 & 6.72 \\
 G4300      &$-$0.69 & 0.40 & 1.48 &   1.2312 & 0.0832 &$-$0.6426 & 0.5499 & 0.2172 & $-$2.51 & 6.69 \\
 H$\gamma_A$&   1.31 & 0.42 & 1.53 &   1.1368 & 0.1181 &   0.9595 & 0.8582 & 0.9515 & $-$10.67& 12.20 \\
 H$\gamma_F$&   0.80 & 0.32 & 1.15 &   1.1245 & 0.0982 &$-$0.2542 & 0.2450 & 0.4774 & $-$3.33 & 8.39 \\
 Fe4383     &   0.85 & 0.36 & 1.31 &   0.5556 & 0.2616 &   2.3222 & 1.2843 & 0.4873 & $-$8.58 & 8.47 \\
 Ca4455     &$-$0.22 & 0.17 & 0.61 &   0.6107 & 0.2103 &   1.1178 & 0.3183 & 0.1305 &    0.42 & 3.72 \\
 Fe4531     &$-$0.33 & 0.09 & 0.31 &   0.7437 & 0.1947 &   1.1023 & 0.6749 & 0.3376 &    0.61 & 6.13 \\
 Fe4668     &$-$0.09 & 0.35 & 1.27 &   0.8227 & 0.0871 &   0.2525 & 0.3734 & 0.4080 & $-$0.88 & 8.72 \\
 H$\beta$   &$-$0.26 & 0.13 & 0.50 &   1.0757 & 0.0511 &   0.1672 & 0.0719 & 0.1347 & $-$0.16 & 7.94 \\
 Fe5015     &$-$0.32 & 0.14 & 0.50 &   0.7301 & 0.0845 &   1.7871 & 0.3799 & 0.2113 &    0.82 & 6.97 \\
 Mg$_1$     &   0.03 & 0.01 & 0.05 &   0.9470 & 0.0279 &$-$0.0077 & 0.0050 & 0.0115 & $-$0.13 & 0.37 \\
 Mg$_2$     &   0.02 & 0.01 & 0.03 &   0.9592 & 0.0123 &$-$0.0091 & 0.0040 & 0.0073 &    0.02 & 0.56 \\
 Mg$b$      &   0.21 & 0.10 & 0.39 &   0.9570 & 0.0505 &$-$0.0258 & 0.1613 & 0.1803 &    0.11 & 17.27 \\
 Fe5270     &$-$0.08 & 0.13 & 0.47 &   0.7173 & 0.2012 &   0.7790 & 0.4942 & 0.3252 &    0.35 & 4.99 \\
 Fe5335     &$-$0.32 & 0.18 & 0.68 &   0.9677 & 0.0881 &   0.3939 & 0.1904 & 0.2338 &    0.77 & 4.77 \\
 Fe5406     &$-$0.27 & 0.15 & 0.53 &   1.0934 & 0.1120 &   0.0038 & 0.1577 & 0.1885 & $-$0.89 & 3.25 \\
 Fe5709     &$-$0.21 & 0.06 & 0.21 &   1.0091 & 0.1178 &   0.0406 & 0.1195 & 0.0674 & $-$1.96 & 1.40 \\
 Fe5782     &$-$0.06 & 0.10 & 0.36 &   0.7652 & 0.1222 &   0.3007 & 0.0719 & 0.0897 & $-$0.20 & 1.27 \\
 NaD        &$-$0.03 & 0.10 & 0.36 &   0.9969 & 0.0272 &   0.1169 & 0.0669 & 0.1350 &    0.30 & 8.57 \\
 TiO$_1$    &   0.01 & 0.01 & 0.02 &   0.7323 & 0.1210 &$-$0.0001 & 0.0029 & 0.0061 & $-$0.00 & 0.58 \\
 TiO$_2$    &$-$0.01 & 0.01 & 0.02 &   0.8658 & 0.0952 &   0.0129 & 0.0034 & 0.0093 & $-$0.02 & 0.94 \\
\enddata
\tablecomments{See Figure~\ref{fig:compGS1x1} for an illustration of the corresponding Lick index correction terms that are being used in Equations~\ref{ln:lincorr} and \ref{ln:eqncorr}. }
\end{deluxetable*}

\end{document}